\documentclass[aps,11pt,prd,longbibliography]{revtex4-1}
\usepackage{tikz}
\usepackage{feynmp-auto}
\usepackage{amsmath}
\usepackage[utf8]{inputenc}
\usepackage[T1]{fontenc} 
\usepackage[utf8]{inputenc}
\usepackage{array}
\usepackage{bbm}
\usepackage{soul}	
\usepackage{xcolor}

\usepackage{epsfig}
\usepackage{color}
\usepackage{slashed}
\usepackage{comment}
\usepackage{epstopdf}
\usepackage{mathrsfs}
\usepackage[euler]{textgreek}
\usepackage{amsmath}
\usepackage{amsthm}
\usepackage{amsfonts}
\usepackage{amssymb}
\usepackage{graphicx}
\graphicspath{ {Figures/} }
\usepackage[font={small}]{caption}
\usepackage{subcaption}
\usepackage{float}
\usepackage{multirow}
\usepackage[hang, flushmargin,bottom]{footmisc} 
\usepackage{hyperref}
\hypersetup{
     colorlinks   = true,
     citecolor    = blue
}
\usepackage{placeins}
\usepackage{enumitem}
\usepackage{slashed}
\usepackage{bbm}
\usepackage{appendix}

\usepackage{titlesec}
\titleformat{\chapter}[display]
  {\normalfont\LARGE\bfseries}
  {\chaptertitlename\ \thechapter}{5pt}{\LARGE}
  \titlespacing*{\chapter}{0pt}{-20pt}{35pt}
\usepackage{bigstrut}
\usepackage{pst-node}
\usepackage{pstricks}
\usepackage{physics}
\usepackage{mathtools}
\usepackage{setspace}
\usepackage{fancyhdr}
\usepackage[makeroom]{cancel}
\usepackage{feyn}
\newcommand{\be}{\begin{equation}}
\newcommand{\ee}{\end{equation}}
\newcommand{\bes}{\begin{equation*}}
\newcommand{\ees}{\end{equation*}}

\usepackage{hyperref}
\hypersetup{%
  colorlinks = true,
  linkcolor  = blue
}
\usepackage{soul}

\newcommand{\beq}{\begin{equation}}
\newcommand{\eeq}{\end{equation}}

\DeclareRobustCommand{\swatch}[1]{\tikz[baseline=-0.6ex]\node[fill=#1,shape=rectangle,draw=black,thick,minimum width=5mm,rounded corners=0.5pt](){};}

\newcommand{\MET}{\ensuremath{E_T^{\rm miss}}}

\definecolor{green}{HTML}{008000}
\definecolor{goldenrod}{HTML}{DAA520}
\definecolor{magenta}{HTML}{FF00FF}
\definecolor{silver}{HTML}{C0C0C0}
\definecolor{indigo}{HTML}{4B0082}
\definecolor{skyblue}{HTML}{87CEEB}
\definecolor{darkgoldenrod}{HTML}{B8860B}
\definecolor{orange}{HTML}{FFA500}
\definecolor{yellow}{HTML}{FFFF00}
\definecolor{saddlebrown}{HTML}{8B4513}
\definecolor{blue}{HTML}{0000FF}
\definecolor{turquoise}{HTML}{40E0D0}
\definecolor{yellow}{HTML}{FFFF00}

\newcommand{\myComment}[1]{}

\begin{document}
\title{\Large{Custodial Symmetry Breaking and Higgs Signatures at the LHC}}
\author{Jon Butterworth$^{1}$, Hridoy Debnath$^{2}$, Pavel Fileviez P{\'e}rez$^{2}$, Francis Mitchell$^{1}$}
\affiliation{
$^{1}$Department of Physics and Astronomy, University College London, Gower St., London, WC1E 6BT, UK \\
$^{2}$Physics Department and Center for Education and Research in Cosmology and Astrophysics (CERCA), Case Western Reserve University, Cleveland, OH 44106, USA }
\date{\today}
\begin{abstract}
We discuss the simplest model that predicts a tree level modification of the $\rho$ parameter from a shift in the $W$-mass without changing the prediction for the $Z$ mass. This model predicts a new neutral Higgs and two charged Higgses, with very similar masses and suppressed couplings to the Standard Model fermions.
We discuss the decay properties of these new scalar bosons, and the main signatures at the Large Hadron Collider. Comparing these signatures for the first time to 
the latest measurements, we show that while masses around 200~GeV are excluded for some scenarios, over a wide range of model parameter space the new bosons can have a mass close to the electroweak scale without violating existing limits from experimental searches or destroying the agreement with measurements. 
We investigate the scenario where the new neutral Higgs is fermiophobic and has a large branching ratio into $W$ gauge bosons and/or photons, and show that this could lead to a signal in the diphoton mass spectrum at low Higgs masses.
We discuss the different signatures that can motivate new measurements and searches at the Large Hadron Collider.
\end{abstract}
\maketitle
\section{INTRODUCTION}
%
The discovery~\cite{ATLAS:2012yve,CMS:2012qbp,CMS:2013btf} of the Higgs boson at the Large Hadron Collider (LHC) was a great step forward for fundamental science. Now, we know how the electroweak symmetry is broken in the Standard Model (SM) of Particle Physics and how all massive elementary particles (except for the neutrinos) can acquire their masses, through the Brout-Englert-Higgs mechanism. The ATLAS and CMS collaborations at the LHC have measured several properties of the Higgs boson, and so far the results are very similar to the predictions in the SM. The exploration of the scalar sector has only started, and one could have other scalar bosons in a more complete theory that could describe physics close to the electroweak scale.

The SM makes a very clear prediction for the relation between the masses of the mediators of the weak force, the $W^\pm$ and $Z$ gauge bosons, i.e. $M_W=M_Z \cos \theta_W$. Here $\cos \theta_W$ is the so-called Weinberg angle. This relation has survived precision tests for almost 40 years. The above relation is a consequence of having only a Higgs $SU(2)_L$ doublet needed for the electroweak symmetry breaking. In the limit when we neglect the hypercharge gauge coupling, all the electroweak gauge bosons have the same mass at tree level because of a residual $SU(2)$ symmetry, the so-called {\textit{custodial symmetry}}~\cite{Sikivie:1980hm,Georgi:1993mps}. If we define the $\rho$ parameter as $\rho=M_W^2/M_Z^2 \cos^2 \theta_W$ one can understand the properties of different extensions of the Standard Model where custodial symmetry could be broken. In the SM the $\rho$ parameter is one at tree level and radiative corrections provide a small deviation. Recently, the CDF collaboration~\cite{CDF:2022hxs} has reported new value for the $W$-mass that, if correct, points towards a new Higgs sector of the theory with $\rho > 1$. Independently of the CDF result it is interesting to investigate different Higgs sectors where custodial symmetry is broken.

In this article, we study the simplest model that predicts an enhancement of the $W$-mass without changing the $Z$-mass at tree level. In this model the Higgs sector is composed of the SM Higgs, $H\sim (1,2,1/2)$, and a real scalar triplet, $\Sigma \sim (1,3,0)$. See Refs.~\cite{Ross:1975fq,Gunion:1989ci,Forshaw:2001xq,Forshaw:2003kh,Chen:2006pb,Chankowski:2006hs,Blank:1997qa,Chivukula:2007koj,FileviezPerez:2008bj,FileviezPerez:2022lxp,Chiang:2020rcv,Lynn:1990zk} for various studies in this context. In this model the custodial symmetry is broken at tree level in the scalar potential. The model predicts three new physical Higgs bosons -- two charged Higges $H^\pm$ and a neutral Higgs $H$ -- with very similar masses. The real triplet does not couple to SM fermions and the new Higgs bosons therefore have suppressed couplings to quarks and leptons. We study in detail the spectrum of the theory, the Higgs decays, and the main production cross sections at the LHC. 

We discuss the most important experimental bounds, which at low masses come from a reinterpretation of stau searches and at intermediate masses come from multi-lepton differential cross-section measurements. We show that the new scalars can be relatively light, with masses close to the electroweak scale. The scenario where the new neutral Higgs is fermiophobic is discussed in detail showing the possibility to have large branching ratios into two photons and into pairs of $W$ gauge bosons.
We point out that in scenarios where the charged Higgses are light, the channels with two taus and missing energy, $\tau^+ \tau^- E_T^{miss}$, are promising to test the predictions of this model. We discuss the decays into gauge bosons such as $H^\pm \to W^\pm Z$ that provide information about the breaking of the custodial symmetry. We provide the first detailed study of all signatures in this model and compare them to LHC data.

The article is organized as follows: In Sec.~\ref{sec:model} we discuss the main properties of the Higgs sector, and the Higgs spectrum. In Sec.~\ref{sec:Higgs bounds} we discuss the bounds from the SM Higgs decays into two photons. In Sec.~\ref{sec:Higgs Decays} we discuss the main properties of the Higgs decays, while in Sec.~\ref{sec:Signatures} we discuss the production cross section, the main signatures and the main collider constraints. Finally, we summarize our main findings in Sec.~\ref{sec:Summary}.
%
\section{THEORETICAL FRAMEWORK}
\label{sec:model}
The simplest model that predicts a tree-level enhancement of the $W$-mass without changing the $Z$-mass contains the SM Higgs,
$H \sim (1,2,1/2)$, and a real triplet, $\Sigma \sim (1,3,0)$. 
We refer to this model as $\Sigma SM$. 
The Lagrangian of the scalar sector is given by
\begin{equation}
\label{eq:lscalar}
\mathcal{L}_\mathrm{scalar} = (D_\mu H)^{\dagger} (D^\mu H) + \Tr (D_\mu \Sigma)^{\dagger}
(D^\mu \Sigma) - V(H,\Sigma),
\end{equation}
where $H^T=( \phi^+,\,\phi^0)$ is the SM Higgs and the real triplet can be written as
\begin{equation}
\Sigma = \frac{1}{2} \left( \begin{array} {cc}
\Sigma^0  &  \sqrt{2} \Sigma^+ \\
\sqrt{2} \Sigma^-  & - \Sigma^0
\end{array} \right),
\end{equation}
with $\Sigma^0$ being real, $\Sigma^+=(\Sigma^-)^*$ and
\begin{eqnarray}
D_\mu\Sigma & = & \partial_\mu\Sigma +ig_2 [W_\mu,\Sigma], \ \text{where} \ 
W_\mu = \sum_{a=1}^3\, W_\mu^a T^a \ .
\end{eqnarray}
Here $W_\mu^a$ and $T^a$ are the gauge bosons and the generators of the $SU(2)_L$ gauge group.
The most general renormalizable scalar potential is given by
\begin{eqnarray}
\label{scalar1}
V(H,\Sigma) & = & - \mu^2  H^\dagger H \ + \ \lambda_0   (H^\dagger H)^2 \ - \ M^2_{\Sigma}  \Tr  \Sigma^2
\ + \ \lambda_1  \Tr  \Sigma^4 \ + \ \lambda_2  (\Tr  \Sigma^2 )^2 \nonumber  \\
& + & \lambda_3  ( H^\dagger H ) \Tr  \Sigma^2 \ + \ \lambda_4  H^\dagger \Sigma^2 H \ + \ a_1  H^\dagger \Sigma H.
\end{eqnarray}
One can write the scalar potential in a more compact form as:
\begin{eqnarray}
V(H,\Sigma) & = & - \mu^2 H^\dagger H \ + \ \lambda_0   \left( H^\dagger H \right)^2 \ - \ \frac{1}{2} M^2_{\Sigma}  F \ + \ \frac{b_4}{4} F^2 \ + \ a_1  H^\dagger \Sigma H  \ + \  \frac{a_2}{2} H^\dag H F,
\label{eq:scalar1b}
\end{eqnarray}
where we use,
\begin{eqnarray}
  F = \left(\Sigma^0\right)^2 + 2\Sigma^+\Sigma^-, \qquad b_4 = \lambda_2+\frac{\lambda_1}{2}, \qquad \text{and} \qquad a_2 = \lambda_3 +\frac{\lambda_4}{2} \ . 
\end{eqnarray}
Notice that the scalar potential of this theory in the limit $a_1 \to 0$ has a $O(4)_H \otimes O(3)_\Sigma$ global symmetry and a discrete symmetry $\Sigma \to - \Sigma$. The $a_1$ parameter determines the breaking of the custodial symmetry and the mass of the new Higgses and their decays. Notice that we can always work with $a_1$ positive because we have the freedom to redefine the triplet field.
The scalar potential is bounded from below when the following conditions are satisfied:
\begin{equation}
\lambda_0 > 0, \ b_4 > 0, \ {\rm{and}}, \ a_2^2 \ < 4 \ \lambda_0 \ b_4.
\end{equation}
Notice that $b_4$ has be non-zero and positive. See Refs.~\cite{Ross:1975fq,Gunion:1989ci,Forshaw:2001xq,Forshaw:2003kh,Chen:2006pb,Chankowski:2006hs,Blank:1997qa,Chivukula:2007koj,FileviezPerez:2008bj,FileviezPerez:2022lxp,Chiang:2020rcv,Ashanujjaman:2023etj} for previous studies in this context. See also the general discussion in Refs.~\cite{Gunion:2002zf,Maniatis:2006fs}.
\subsection{Mass Spectrum}
The fields, $H$ and $\Sigma$, can be written as
\begin{eqnarray}
\label{eq:shifted}
H  & =&
\left(\begin{array}{c} \phi^+ \\ (v_0+h^0+i\xi^0) /\sqrt{2} \end{array}\right), \qquad \text{and} \ \qquad \
\Sigma  =  \frac{1}{2} \left( \begin{array}{cc}
x_0+\sigma^0  & \sqrt{2}\Sigma^+ \\
\sqrt{2}\Sigma^-  & - x_0-\sigma^0
\end{array} \right) \ ,
\end{eqnarray}
where $v_0$ and $x_0$ are the SM Higgs and triplet scalar vevs, respectively.

Minimizing the tree-level scalar potential one finds
\begin{eqnarray}
\left( - \mu^2 +  \lambda_0  v_0^2  -  \frac{a_1  x_0 }{2} +   \frac{a_2  x_0^2 }{2} \right)  v_0 & = & 0  ,
\label{mina}
\\
- M_{\Sigma}^2  x_0  +  b_4 x_0^3  -  \frac{a_1  v_0^2}{4 } + \frac{a_2  v_0^2  x_0}{2} & = & 0  ,
\label{minb}
\end{eqnarray}
and
\begin{eqnarray}
\label{b4}
b_4  &>&  \frac{1}{8 x_0^2} \left( - \frac{a_1 v_0^2}{x_0}  +  \frac{\left(- a_1 + 2 a_2 x_0 \right)^2}{2 \lambda_0} \right).
\end{eqnarray}
Notice that when the parameter $a_1=0$ the triplet vacuum expectation value could be non-zero, $x_0 \neq 0$, but in this case the global $O(3)_\Sigma$ is broken and one has extra charged Goldstone bosons. Therefore, this case is ruled out by experiment. 
The only realistic scenario when $x_0 \neq 0$ corresponds to the case when $a_1$ is different from zero, i.e.
\begin{equation}
x_0 \simeq \frac{a_1 v_0^2}{4 (-M_\Sigma^2 + a_2 v_0^2/2)},
\label{x0a2}
\end{equation}
when $v_0 \gg x_0$. In the above equation we have neglected the term $b_4 x_0^3$ because it is much smaller than the other terms in Eq.(\ref{minb}). We have discussed above that $a_1$ is positive and $M_{\Sigma}^2 > 0$ when the vacuum with $x_0 \neq 0$ and $v_0 \neq 0$ is a global minimum, see a detailed discussion in Appendix~\ref{appendix2}. Notice that from Eq.(\ref{x0a2}) one can see that $a_2$ must be positive ($a_2 > 2 M_\Sigma^2/v_0^2$) because $x_0$, $a_1$ and $M_\Sigma^2$ are positive. Therefore, in the rest of the paper we discuss the phenomenology of this model when $a_2$ is positive. It is important to mention that the charged Higgs mass given in Eq.(\ref{chargedmass}) implies that $x_0$ must be positive because $a_1 > 0$ and $M_{H^\pm}^2 > 0$.

Using the above conditions we find the mass matrices of the neutral Higgses in the basis, ($h^0$, $\sigma^0$), and charged Higgses in the basis, ($\phi^\pm$, $\Sigma^\pm$). These mass matrices are given by 
\begin{equation}
\label{eq:massmtrx}
{\cal M}_{0}^2 = \left( \begin{array} {cc}
2 \lambda_0 v_0^2  &  - a_1 v_0 / 2 + a_2  v_0  x_0 \\
- a_1  v_0/ 2  +  a_2  v_0  x_0  & 2 b_4  x_0^2 + \frac{a_1  v_0^2}{4  x_0}
\end{array} \right),
\end{equation}
and
\begin{equation}
{\cal M}_{\pm}^2 = \left( \begin{array} {cc}
a_1 x_0  &  a_1 v_0 / 2  \\
a_1  v_0/ 2 & \frac{a_1 v_0^2}{4 x_0}
\end{array} \right).
\end{equation}
In our convention, the physical mass eigenstates are defined by
\begin{eqnarray}
\label{eq:neutralmix}
\left( \begin{array}{c} h \\ H\end{array} \right) & = & \left(\begin{array}{ccc} \cos \theta_0 & \sin \theta_0 \\ - \sin \theta_0 & \cos \theta_0\end{array} \right) \left( \begin{array}{c} h^0 \\ \sigma^0 \end{array} \right) \  ,\\
& & \nonumber \\
\label{eq:chargemix}
\left( \begin{array}{c} H^\pm \\ G^\pm \end{array} \right) & = & \left(\begin{array}{cc} -\sin \theta_\pm & \cos \theta_\pm  \\ \cos\theta_\pm & \sin \theta_\pm  \end{array} \right) \left( \begin{array}{c} \phi^\pm \\ \Sigma^\pm \end{array} \right) \ .
\end{eqnarray}
The eigenvalues of these matrices are the tree-level masses of the physical scalars ($h$, $H$, $H^\pm$) of the theory, and are given by
\begin{eqnarray}
\label{mneut1}
M_{h}^2 & = & \lambda_0 v_0^2 \left(1+ \frac{1}{\cos 2 \theta_0}\right) + \left( \frac{a_1 v_0^2}{8x_0} + b_4 x_0^2\right)\left(1-\frac{1}{\cos 2 \theta_0}\right)\,, \\
M_{H}^2 & = & \lambda_0 v_0^2 \left(1-\frac{1}{\cos 2 \theta_0} \right) + \left(\frac{a_1 v_0^2}{8x_0} + b_4 x_0^2\right)\left(1+\frac{1}{\cos 2\theta_0} \right), 
\end{eqnarray}
and
\begin{eqnarray}
M_{H^{\pm}}^2 &=&  a_1 x_0 \left( 1 + \frac{ v_0^2}{4 x_0^2} \right),
\label{chargedmass}
\end{eqnarray}
where $\theta_0$ is a mixing angle defined in Eq.~(\ref{eq:neutralmix}). 
From Eq.~(\ref{chargedmass}) one can see that $x_0$ is always positive because $a_1$ is positive. 
Notice that since $v_0\gg x_0$ the new Higgses have very similar masses, i.e. $M_{H^\pm}\approx M_H$.
The mixing angles are given by
\begin{equation}
\label{eq:thetazero}
\tan 2 \theta_0 =  \frac{4 v_0 x_0(-a_1 + 2 x_0 a_2)}{8 \lambda_0 v_0^2 x_0 - 8
b_4 x_0^3 - a_1 v_0^2},\hspace{5mm}\text{and}\hspace{5mm}\tan 2 \theta_+ =  \frac{4 v_0 x_0}{4 x_0^2-v_0^2}\,.
\end{equation}
\begin{figure} [t]  
    \begin{subfigure}[h]{0.45\textwidth}
        \centering
         \includegraphics[width=\textwidth]{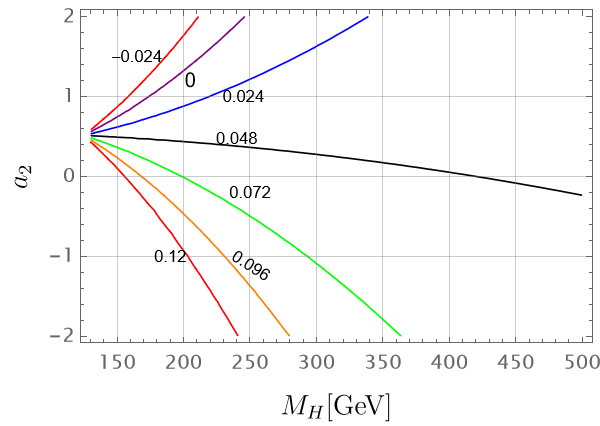}
         \caption{\label{fig:Fig1a}}
    \end{subfigure}
     \hfill
     \begin{subfigure}[h]{0.49\textwidth}
         \centering
         \includegraphics[width=\textwidth]{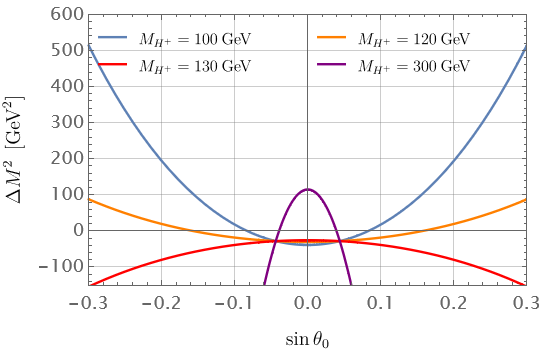}
         \caption{\label{fig:Fig1b}}
     \end{subfigure}
       \caption{(a) Neutral Higgs mixing angle, $\theta_0$ for different values of $a_2$ and $M_H$. (b) $\delta M $ as a function of $\sin{\theta_0}$. Here we used $x_0 = 5.4$ GeV and $b_4 =1$. }
       \label{fig:Fig1}
 \end{figure}
It is very important to mention that the $W$ and $Z$ gauge boson masses are given by
\begin{eqnarray}
M_W^2&=&\frac{g_2^2}{4} (v_0^2 + 4 x_0^2), \ {\rm and} \
M_Z^2=\frac{(g_1^2+g_2^2)}{4} v_0^2.
\end{eqnarray}
Notice that only the $W$-mass is modified at tree level.
This simple model has the following free parameters:
\begin{equation}
x_0, \ M_{H^\pm}\approx M_H, \ a_2,  {\rm{and}} \ b_4.
\end{equation}
In Ref.~\cite{FileviezPerez:2022lxp} the possibility of explaining the recent CDF result for the $W$-mass was discussed, and the authors found that $x_0\approx 5.4$ GeV. See also Ref.~\cite{FileviezPerez:2008bj} for a detailed discussion of this model where the authors have shown that the one-loop contributions to the $S$, $T$ and $U$ parameters are very small. For a recent study see the results in Ref.~\cite{Cheng:2022hbo}. For the bounds coming from perturbative unitarity see the discussion in Ref.~\cite{Chivukula:2007koj}.

In Fig.~\ref{fig:Fig1a} we show the values of the neutral mixing angle for different values of the Higgs mass and the $a_2$ parameter. Since $\theta_0$ is always very small, the couplings of the new neutral Higgs to fermions are suppressed. The mixing angle in the charged sector given in Eq.(\ref{eq:thetazero}) is also very small and the couplings of the new charged Higgses to the SM fermions are also suppressed~\footnote{See the appendix for the Feynman rules.}.

To understand the Higgs spectrum one can define the parameter $\Delta M ^2 = M_{H^+}^2-M_H^2$. In Fig.~\ref{fig:Fig1b} we show numerical results for the mass splitting. In the small mixing angle region, the mass splitting is too small compared to the Higgs masses to have a significant impact on the phenomenology. For example, when $\sin{\theta_0}=0.05$ and $M_{H^+}=100$ GeV, the mass splitting between the Higgses is 
$\delta M = M_{H}-M_{H^+} = 124 \hspace{0.1 cm}\text{MeV}$.
Given this, we will assume $M_{H^\pm} = M_H$ for the rest of the paper. 
\subsection{Bounds from $h \to \gamma \gamma $}
\label{sec:Higgs bounds}
\begin{figure}[t]   
         \centering
         \includegraphics[width=0.6\textwidth]{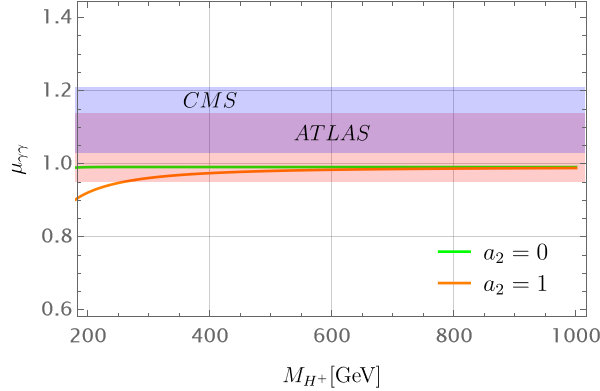}
       \caption{$\mu_{\gamma \gamma}$ for different values of $M_{H} $ and $a_2$. Here the light red and blue shaded regions show the bounds on $\mu_{\gamma \gamma}$ from ATLAS and CMS experiments, respectively.}
\label{fig:mugg}
 \end{figure}
The existence of new charged Higgses could change the properties of the SM Higgs. In particular, the decay of the SM-like Higgs into two photons could be modified because the charged Higgs can enter into the effective coupling at one-loop level. We use the following relation to constrain the interaction between the SM-like Higgs and the charged Higgses: 
\begin{equation}
    \mu_{\gamma \gamma} = \frac{\sigma (pp \to h) \times {\rm BR}(h\to \gamma \gamma)}{\sigma (pp \to h)_{\rm SM} \times {\rm BR}(h\to \gamma \gamma)_{\rm SM}}.
    \label{mugammagamma}
\end{equation}
The ATLAS collaboration measured this ratio and find that $\mu_{\gamma \gamma}=1.04^{+ 0.10}_{-0.09}$~\cite{ATLAS:2022tnm} , while CMS finds $\mu_{\gamma \gamma}=1.12 \pm 0.09$~\cite{CMS:2021kom}. Using these results we show the predictions for $\mu_{\gamma \gamma}$ in our model in Fig.~\ref{fig:mugg} for different values of $a_2$ and the Higgs mass. It can be seen that when $a_2$ is zero or positive the LHC bounds can be satisfied in most of the parameter space. If the Higgs mass is below approximately 250 GeV one cannot satisfy the bounds when $a_2=1$. Therefore, these bounds have significant impact on the allowed parameter space, essential ruling out $a_2 > 0$ when the new charged Higgs is very light.
%
%
\section{DECAYS OF THE NEW HIGGS BOSONS}
\label{sec:Higgs Decays}
\subsection{Charged Higgses}
\begin{figure}[t]   
    \begin{subfigure}[h]{0.49\textwidth}
        \centering
         \includegraphics[width=\textwidth]{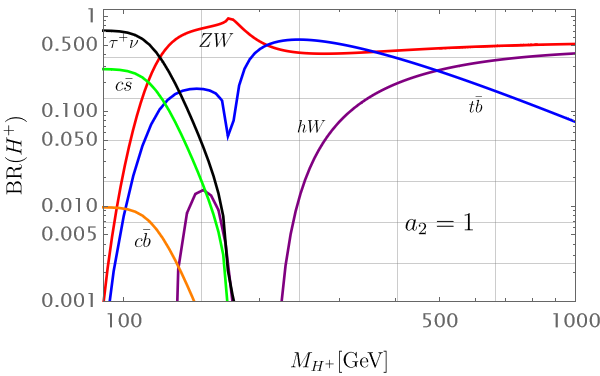}
         \caption{\label{fig:ChargedHiggses-1}}
     \end{subfigure}
     \hfill
    \begin{subfigure}[h]{0.49\textwidth}
        \centering
         \includegraphics[width=\textwidth]{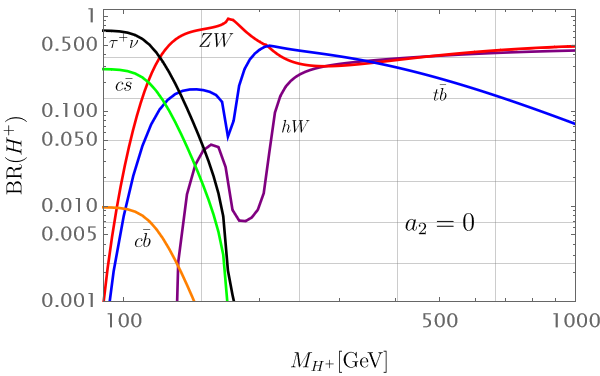}
         \caption{\label{fig:ChargedHiggses-0}}
     \end{subfigure}
      \caption{Charged Higgs decays for different values of $a_2$.}
 \label{fig:ChargedHiggses}
 \end{figure}

The main decays of the charged Higgses are
\begin{equation}
H^+ \to W^+ Z, \ h W^+, \ t \bar{b}, \ \tau^+ \nu, \ c \bar{s}.
\end{equation}
where the decay $H^+ \to W^+ Z$ is a direct consequence of custodial symmetry breaking. In Fig.~\ref{fig:ChargedHiggses} we show the branching ratios for the decays of the charged Higgses for different values of the Higgs mass and $a_2$ parameter. In the scenarios where the charged Higgses are light, the largest branching ratio is for the decay $H^\pm \to \tau^\pm \nu$, while in the heavy region, the main decays are $H^+ \to t \bar{b}$, $H^\pm \to W^\pm Z$ and $H^\pm \to h W^\pm$. Notice that when the Higgs mass is around 150 GeV the main decay is $H^\pm\to W^\pm Z$ independent of the value of $a_2$. Here we assume the value, $x_0=5.4$ GeV, for the triplet vacuum expectation value, although our
calculations show that there is no significant dependence of the branching ratios on $x_0$ over the parameter range considered.
With $a_2=0$, the most important decay is $H^+ \to \tau^+ \nu$ when $M_{H^+}=95$ GeV, when $M_{H^+}=150$ GeV the charged Higgs decays mainly into $WZ$, and when $M_{H^+}=300$ GeV the decay into the third family of quarks dominates. These three scenarios tell us about the main features of the charged Higgs decays when we change the Higgs mass.
%
\subsection{Neutral Higgs}
\label{neutralhiggs}
\begin{figure}[t]   
    \begin{subfigure}[h]{0.49\textwidth}
        \centering
         \includegraphics[width=\textwidth]{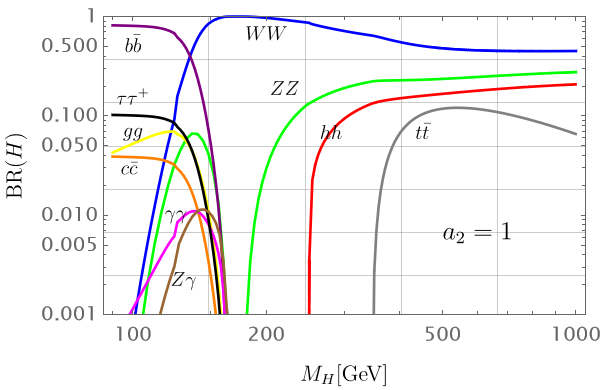}
         \caption{\label{fig:Neutral-Higgs-decays-1}}
     \end{subfigure}
     \hfill
     \begin{subfigure}[h]{0.49\textwidth}
         \centering
         \includegraphics[width=\textwidth]{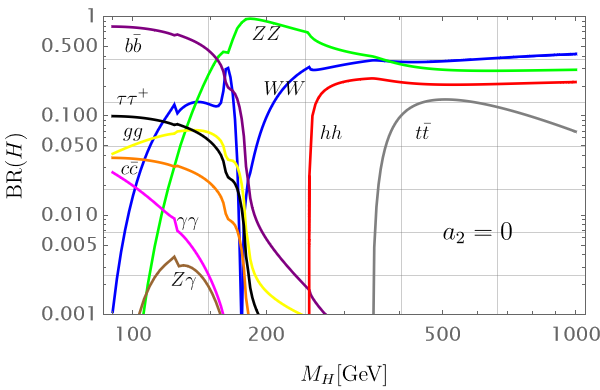}
         \caption{\label{fig:Neutral-Higgs-decays-0} }
     \end{subfigure}
       \caption{CP-even Higgs decays for different values of $a_2$ when $x_0=5.4$ GeV.}
\label{fig:Neutral-Higgs-decays}
     \end{figure}

The new CP-even Higgs decays mainly into the gauge bosons, the SM-like Higgs, and the third generation of quarks:
\begin{equation}
H \to WW, ZZ, b\bar{b}, t \bar{t}, h h.
\end{equation}
In Fig.~\ref{fig:Neutral-Higgs-decays} we show these decays for different values of the Higgs mass and the $a_2$ parameter. Notice that changing $a_2$ from $1$ to $0$ the decays to gauge bosons change dramatically, because the $WW$ channel is suppressed when the Higgs mass is around $250$ GeV. There are three main scenarios as in the case of the charged Higgs decays: a) when the Higgs mass is around 100 GeV the main decay is into two bottom quarks, b) in the intermediate region the $WW$ channel dominates when $a_2=1$ or the $ZZ$ channel when $a_2=0$ and
c) in the heavy region the decays into gauge bosons and into the SM Higgs are similar.
As with the charged Higgs, the branching ratios have very little dependence on the triplet vacuum expectation value, $x_0$. Notice that when the Higgs mass is around $150$ GeV one could have access to most of the decay channels because many of them have large branching ratios.
%
\subsubsection{Fermiophobic Scenario}
\label{sec:fermiophobic}

\begin{figure}[t]   
  \begin{subfigure}[h]{0.49\textwidth}
    \centering
    \includegraphics[width=\textwidth]{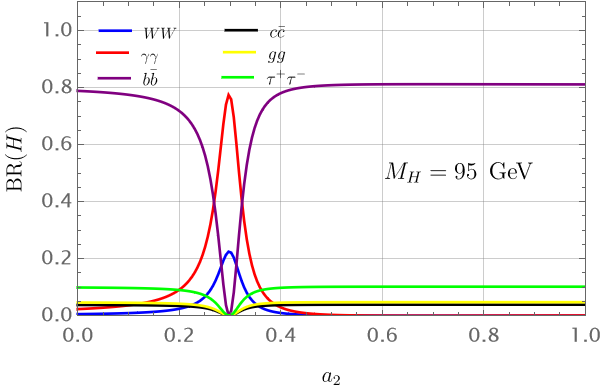}
    \caption{\label{fig:Neutral-Higgs-decays-a2-95} }
  \end{subfigure}
     \hfill
     \begin{subfigure}[h]{0.49\textwidth}
         \centering
         \includegraphics[width=\textwidth]{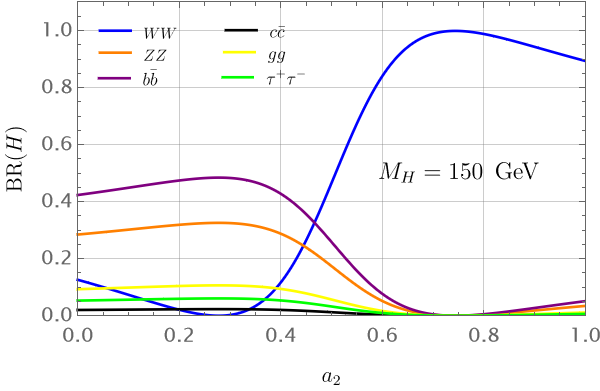}
         \caption{\label{fig:Neutral-Higgs-decays-a2-150} }
     \end{subfigure}
     \hfill
       \caption{CP-even Higgs decays for different values of $M_H$ when $x_0=5.4$ GeV.}
\label{Neutral-Higgs-decays-a2}
     \end{figure}

In this model only the new neutral Higgs could be fermiophobic when $\theta_0$ goes to zero. This limit can be achieved when $a_1 \to 2 x_0 a_2$. This condition is consistent with the minimization condition in Eq.~(\ref{minb}).
Using Eq.~(\ref{eq:thetazero}) and the perturbative bound, $a_2 \leq 2 \sqrt{\pi}$, one finds
\begin{equation}
M_H \simeq M_{H^\pm} \lesssim 330 \ {\rm GeV}.    
\end{equation}
This condition holds only in the fermiophobic limit.
Since the neutral Higgs couplings to fermions goes to zero one can have a large branching ratio into $WW$ and into two photons. In Fig.~\ref{Neutral-Higgs-decays-a2} we show the branching ratios of the neutral Higgs for different values of $a_2$ and for two different values of the Higgs mass, $M_H=95$ GeV (left-panel) and $M_H=150$ GeV (right-panel). Notice that when the Higgs mass is $95$ GeV the fermiophobic limit is realized when $a_2$ is around 0.3, while in the case of $M_H=150$ GeV, the neutral mixing angle goes to zero when $a_2 \sim 0.75 $. In this regime, the bound from $h \to \gamma \gamma$, see Eq.~(\ref{mugammagamma}), is important. For example, when $M_H=95$ GeV and $a_2=0.1$ one finds that $\mu_{\gamma \gamma}$ is around 0.94, while for $M_H=150$ GeV and $a_2 = 0.5$ one has $\mu_{\gamma \gamma}$ around 0.93. Both scenarios are very close to the experimental limis on $\mu_{\gamma \gamma}$.
In the left-panel one can have large branching ratios into $WW$ and $\gamma \gamma$ in the fermiophobic limit, while in the right-panel the $WW$ channel dominates. For other studies with fermiophobic Higgses see Refs.~\cite{Akeroyd:2005pr,Delgado:2016arn,Akeroyd:2003xi}.
\section{SIGNATURES AT THE LHC}
\label{sec:Signatures}
\begin{figure}[t]
     \centering 
         \includegraphics[width=0.6\textwidth]{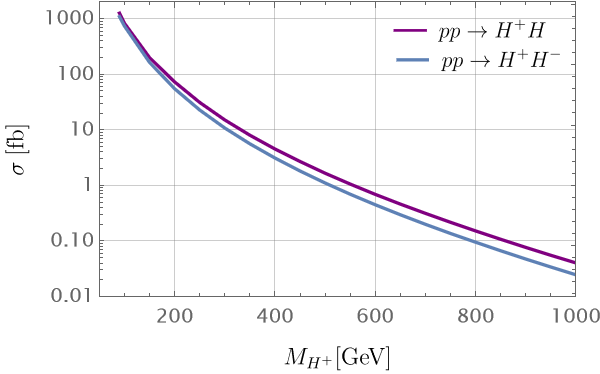}
    \caption{Production cross section  for $pp \to Z^*, \gamma \to H^+ H^-$ and $pp \to (W^{\pm})^* \to H^{\pm} H$ for different values of $M_H$ at $\sqrt{s}=13 $ TeV. Here we assumed $M_H = M_{H^+}$ and used MadGraph5~\cite{Alwall_2011} to compute the cross sections.}
    \label{fig:Drell-Yan}
\end{figure}

The new Higgs bosons predicted in this theory have suppressed couplings to the top and bottom quarks. The gluon fusion production channel is therefore very suppressed for the new Higgs bosons. The vector boson fusion production mechanisms are also very suppressed due to the small vev~\cite{FileviezPerez:2008bj}. This leaves Drell-Yan processes as the dominant production mechanism:
\begin{equation}
pp \to Z^*, \gamma \to H^+ H^-, \  {\textrm{and}} \ pp \to (W^{\pm})^* \to H^{\pm} H.
\end{equation}
The cross section for 13~TeV proton-proton collisions is shown in Fig.~\ref{fig:Drell-Yan}. Notice that only when the Higgs mass is below $500$ GeV is the cross section above $1$ fb. Therefore, one cannot expect a large number of events when the Higgs mass is above $500$ GeV.
One interesting aspect of this theory is that the charged Higgses can be very light yet remain in agreement with all constraints because the couplings to the SM fermions are so suppressed by the mixing angles between the Higgses. 
As expected, in the light Higgs scenario, most events correspond to the decays $H^\pm \to \tau^\pm \nu$ and $H \to b \bar{b}$. In particular, the signatures with two taus and two neutrinos,
$\tau^\pm \tau^\mp E_T^{miss}$, can be used to look for these light charged Higgses. We will discuss the main collider bounds for these signatures, as well as those at intermediate masses. 
%
%
\subsection{Reinterpretation of Searches}
Since the charged Higgses could in principle have mass around the electroweak scale, we would like to find a lower bound on their masses using experimental constraints.
This model evades many constraints from direct searches for charged and exotic neutral Higgs bosons.
The coupling to the top is low enough that limits on the charged Higgs
from top decays~\cite{CMS:2012fgz,ATLAS:2015zgd}, generally at the few percent level, have no impact.
The recent ATLAS search for charged Higgs bosons~\cite{ATLAS:2022zuc} sets limits in the Georgi–Machacek model on the production cross section of $H^\pm$ at around 100~fb for masses in the range 200-400~GeV and a branching to $WZ$ of unity, which is above the expectation shown in Fig.~\ref{fig:Drell-Yan}. 
Limits on neutral Higgs production, for example \cite{ATLAS:2020tlo}, are in a similar range.
As we will show, the new charged and neutral Higgses in this model can have masses around the electroweak scale and avoid the bounds mentioned above.

\begin{figure}[t]
         \centering
         \includegraphics[width=0.6\textwidth]{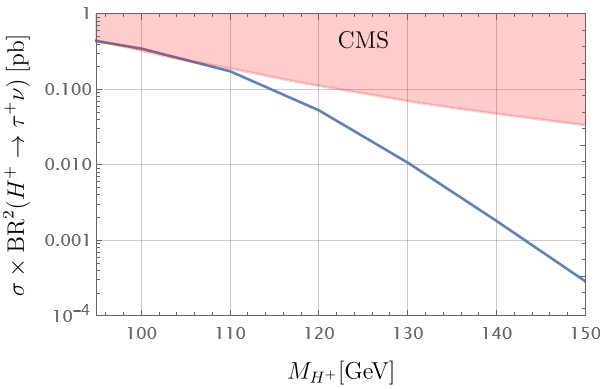}
    \caption{CMS bounds from stau searches~\cite{CMS:2022syk} as a function of the new charged Higgs mass using $a_2=1$ and $x_0=5.4$ GeV. These bounds are not sensitive to the value of $a_2$ and $x_0$ in the low mass region.}
    \label{fig:CMS}
\end{figure}

One can use the searches for supersymmetric particles to set a bound on the mass of the charged Higgses using the channel $\tau^\pm \tau^\mp E_T^{miss}$. In Fig.~\ref{fig:CMS} we show the predictions for $\sigma (pp \to H^+ H^-) \times {\rm{BR}}^2(H^+ \to \tau^+ \nu)$ (blue line) for different values of the Higgs mass. The region in red is the region excluded by CMS if we use the stau searches for the channel $pp \to \tilde{\tau}^+ \tilde{\tau}^- \to \tau^+ \tau^- \tilde{\chi}_1^0 \tilde{\chi}_1^0$~\cite{CMS:2022syk}. Here $\tilde{\tau}^\pm$ is the supersymmetric partner of the tau lepton and $\tilde{\chi}_1^0$ is the lightest supersymmetric particle, i.e. the lightest neutralino.
Only in the region below $110$ GeV does the exclusion approach the predicted signal.
This analysis is very naive, a dedicated search, or an optimised measurement of ditau production, would very likely improve the sensitivity to this scenario.
It is interesting to note that the predicted cross-section-times-branching-ratio for neutral Higgs decay into two photons at low Higgs masses
can be large in the fermiophobic scenario, see Figures (\ref{Neutral-Higgs-decays-a2}) and (\ref{fig:Drell-Yan}), and we will discuss the relevant limits below.
%
\subsection{Differential Cross Section Measurements}
%
\begin{figure}[h]
  \centering
    \begin{subfigure}[h]{0.45\textwidth}
    \centering
    \includegraphics[width=\textwidth]{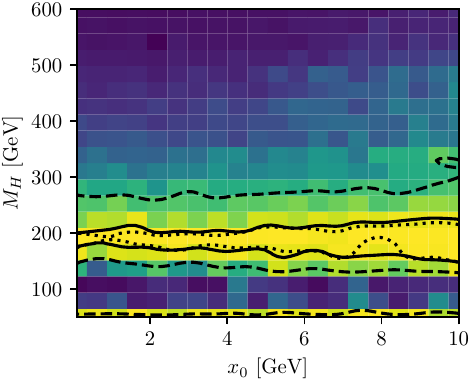}
    \caption{$a_2=0, b_4=10^{-4}$ }
    \end{subfigure}
    \begin{subfigure}[h]{0.45\textwidth}
    \centering
    \includegraphics[width=\textwidth]{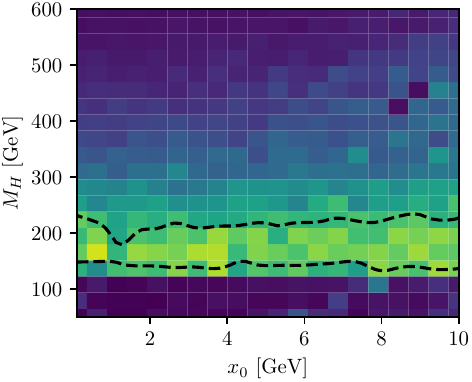}
    \caption{$a_2=1, b_4=10^{-4}$}
    \end{subfigure}
    \begin{subfigure}[h]{0.08\textwidth}
    \centering
    \includegraphics[width=\textwidth]{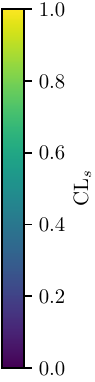}
    \caption*{}
    \end{subfigure}
    \begin{subfigure}[h]{0.45\textwidth}
    \centering
    \includegraphics[width=\textwidth]{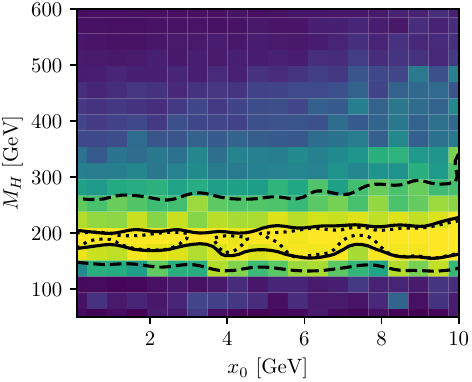}
    \caption{$a_2=0, b_4=2\sqrt{\pi}$ }
    \end{subfigure}
    \begin{subfigure}[h]{0.45\textwidth}
    \centering
    \includegraphics[width=\textwidth]{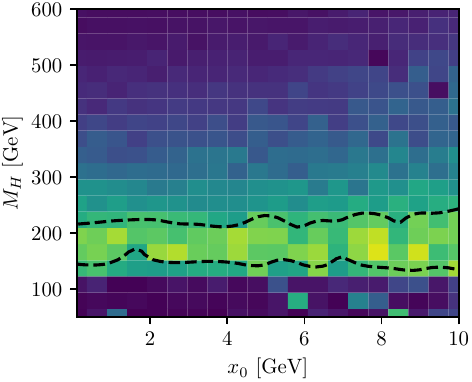}
    \caption{$a_2=1, b_4=2\sqrt{\pi}$}
    \end{subfigure}
    \begin{subfigure}[h]{0.08\textwidth}
    \centering
    \includegraphics[width=\textwidth]{combinedMeshcbar.pdf}
    \caption*{}
    \end{subfigure}
  \caption[]{Exclusions in the plane of vev ($x_0$) vs Higgs mass ($M_H$).
    The dotted line indicates the expected 95\% exclusion, the solid line the actual 95\% exclusion, and the dashed line the actual 68\% exclusion. 
    \label{fig:scans}}
\end{figure}
\begin{figure}[h]
  \centering
  \subfloat{\includegraphics[scale=0.6]{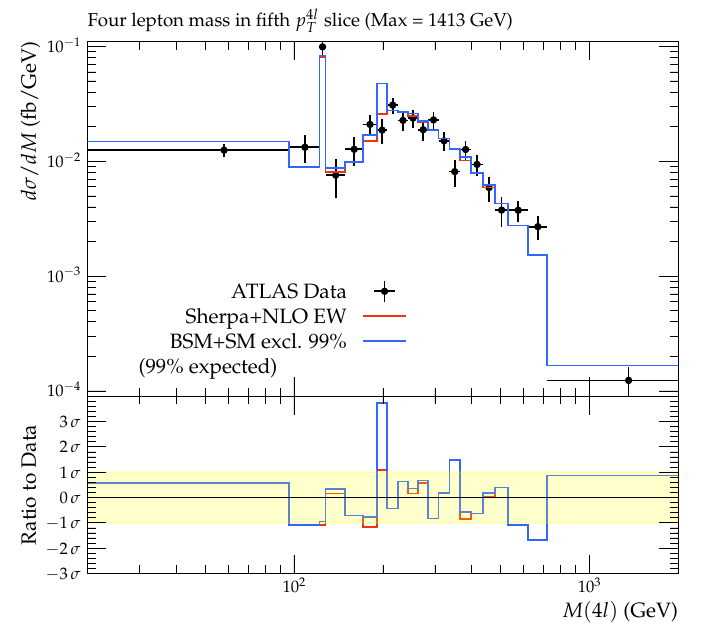}}
  \caption[]{The four-lepton invariant mass distribution for $a_2 = 0, M_H = 195$~GeV, and $x_0 = 5.4$~GeV, compared to the ATLAS
    measurement~\cite{ATLAS:2021kog}. The Standard Model prediction from the experimental paper, made using Sherpa~\cite{Sherpa:2019gpd}, is also shown. The $p_T$ of the four-lepton system is in the range $100$~GeV~$ < p_T(4\ell) < 600$~GeV.
    \label{fig:m4l}}
\end{figure}
\begin{figure}
    \begin{subfigure}[h]{0.6\textwidth}
    \centering
    \includegraphics[width=\textwidth]{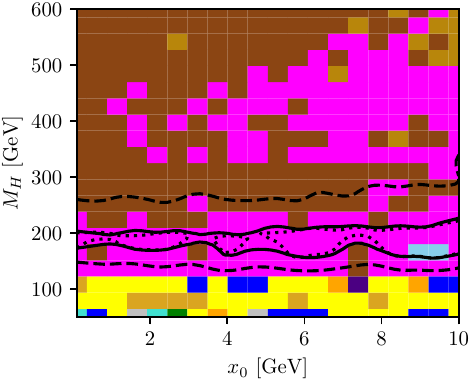}
    \caption{$a_2 = 0$}
    \end{subfigure}
    \begin{subfigure}[h]{0.6\textwidth}
    \centering
    \includegraphics[width=\textwidth]{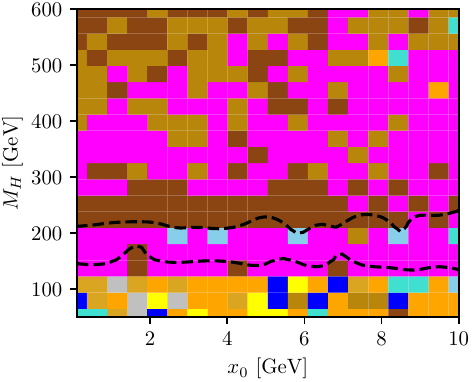}
    \caption{$a_2 = 1$}
    \end{subfigure}
    \caption[]{The most sensitive final states in the $(x_0, M_H)$ parameter plane. The citations
    indicate measurements which provide an individual exclusion above 68\%.
    \label{fig:pools} }
    \begin{tabular}{llll}      
        \swatch{magenta}~4$\ell$ \cite{ATLAS:2021kog} & 
        \swatch{skyblue}~$\ell_1\ell_2$+\MET{} \cite{ATLAS:2019rob} & 
        \swatch{turquoise}~$\ell_1\ell_2$+\MET{}+jet  & 
        \swatch{indigo}~$\ell^+\ell^-$+\MET{}  \\ 
        \swatch{saddlebrown}~hadronic $t\bar{t}$  & 
        \swatch{goldenrod}~$\gamma$+\MET{}  & 
        \swatch{silver}~jets  &
        \swatch{orange}~$\ell^+\ell^-$+jet  \\
        \swatch{blue}~$\ell$+\MET{}+jet  & 
        \swatch{darkgoldenrod}~$\ell^+\ell^-\gamma$  & 
        \swatch{yellow}~$\gamma$ and $\gamma\gamma$ & 
        \swatch{green}~\MET{}+jet  \\ 
\end{tabular}
\end{figure}
\begin{figure}
  \centering
    \begin{subfigure}[h]{0.6\textwidth}
    \centering
    \includegraphics[width=\textwidth]{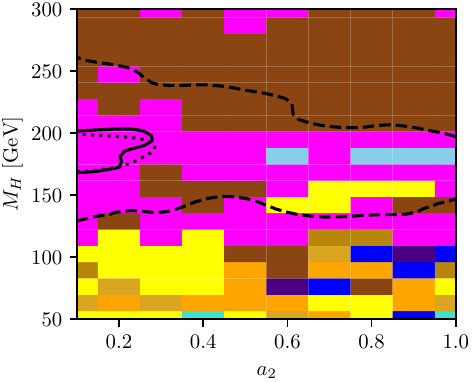}
    \caption{}
    \end{subfigure}
    \caption[]{The most sensitive final states in the $(a_2, M_H)$ plane for $x_0 = 5.4$~GeV. The citations
    indicate measurements which provide  an individual exclusion above 68\%.}
    \label{fig:pools_a2} 
    \begin{tabular}{llll}
        \swatch{yellow}~$\gamma$ and $\gamma\gamma$ \cite{ATLAS:2017lpx,ATLAS:2021mbt,ATLAS:2014yga} &       
        \swatch{magenta}~4$\ell$ \cite{ATLAS:2021kog} & 
        \swatch{darkgoldenrod}~$\ell^+\ell^-\gamma$ \cite{ATLAS:2022wmu} & 
        \swatch{goldenrod}~$\gamma$+\MET{} \cite{ATLAS:2016qjc,ATLAS:2018nci} \\ 
        \swatch{turquoise}~$\ell_1\ell_2$+\MET{}+jet & 
        \swatch{skyblue}~$\ell_1\ell_2$+\MET{} & 
        \swatch{indigo}~$\ell^+\ell^-$+\MET{} & 
        \swatch{orange}~$\ell^+\ell^-$+jet  \\
        \swatch{blue}~$\ell$+\MET{}+jet  & 
        \swatch{saddlebrown}~hadronic $t\bar{t}$ &
\end{tabular}
\end{figure}
\begin{figure}
\begin{subfigure}[h]{0.5\textwidth}
    \centering
    \includegraphics[width=\textwidth]{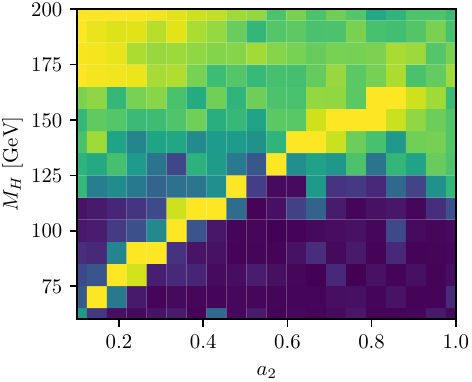}
    \end{subfigure}
   \begin{subfigure}[h]{0.11\textwidth}
    \centering
   \includegraphics[width=\textwidth]{combinedMeshcbar.pdf}
    \end{subfigure}
     \caption{A higher-granularity scan highlighting the fermiophobic case.}
     \label{fig:pools_a2-high}
    \end{figure}
Given the wide variety of decays which appear for different Higgs masses, several potentially relevant final states have been measured at the LHC, and are available in the Rivet~\cite{Bierlich:2019rhm} library. We use the Contur~\cite{Butterworth:2016sqg,Buckley:2021neu} toolkit to perform a signal-injection of events predicted by the $\Sigma SM$ model on these measurements, to see whether such a signal would have been visible. 

To do this, we have implemented the model in Feynrules~\cite{Alloul:2013bka} and exported as a UFO directory~\cite{Degrande:2011ua}. For each model parameter point, 30,000 events are generated using Herwig~\cite{Bellm:2015jjp} via its UFO interface. These events are then passed through Rivet, which applies the fiducial, particle level selections appropriate to each measurement, determining how many BSM events would have entered each measured cross section. Contur then uses a $\chi^2$ test to evaluate the likelihood ratio in the $CL_s$ method, 
which is interpreted as a potential exclusion probability for the model parameter point concerned, using the relevant SM background prediction.
\begin{figure}[h]
  \centering
  \subfloat{\includegraphics[scale=0.6]{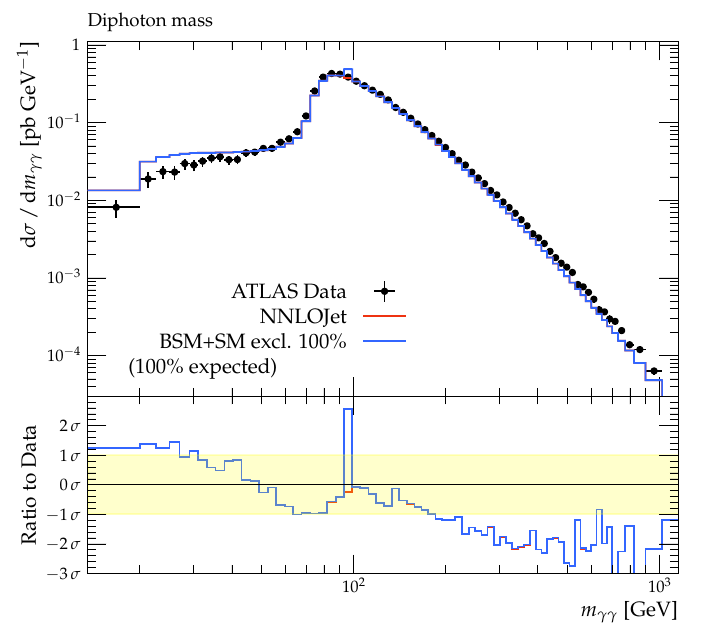}}
  \caption[]{The diphoton invariant mass distribution for $a_2 = 0.3, M_H = 95$~GeV, and $x_0 = 5.4$~GeV, compared to the ATLAS
    measurement~\cite{ATLAS:2021mbt}.
    The Standard Model prediction from the experimental paper, made using NNLOJet~\cite{Gehrmann:2018szu}, is also shown.
    \label{fig:diphoton}}
\end{figure}
Contur contains a very wide range of differential cross section measurements (several hundred measurements in all), covering in particular several diboson and top measurements from ATLAS and CMS which could potentially receive contributions from the model; although we note that no measurements involving taus are currently available in this framework.

Following the discussion in the previous sections, we perform parameter scans in the $(x_0, M_H)$ plane, for $a_2 = 0$ and $1$ and $b_4 = 10^{-4}$ and $2\sqrt{\pi}$, as a representative sampling of the valid parameter space. The results are shown in Fig.~\ref{fig:scans}. It can be seen that much of the parameter space is essentially unconstrained by the measurements considered, apart from a band close to $M_H = 200$~GeV when $a_2=0$. This band expands slightly as $x_0$ increases. 
The measurement mostly responsible for this exclusion is the ATLAS inclusive four-lepton measurement~\cite{ATLAS:2021kog}. The signature comes from the neutral heavy Higgs decaying to $ZZ$, a branching fraction which peaks for $M_H \approx 2M_Z$ when $a_2 = 0$ (see Fig.~\ref{fig:Neutral-Higgs-decays-0}) and leads to a narrow peak in the four-lepton invariant mass distribution, as shown in Fig.~\ref{fig:m4l}. This peak is most visible when the four-lepton system has high $p_T$, due to the recoil against the charged Higgs.
In the $b_4 = 10^{-4}$ case, there is also an exclusion at the lowest masses. Since the value of $b_4$ has little influence over the parameter region of interest, from here on we will fix it at $2\sqrt{\pi}$.

Fig.~\ref{fig:pools} indicates the most sensitive final states for two different $a_2$ scenarios. 
Other measurements do also contribute to the exclusion, and in particular at low mass the ATLAS diphoton measurement~\cite{ATLAS:2021mbt} excludes the model for $a_2 = 0$ at the lowest Higgs masses for low values of $b_4$ (see Fig.~\ref{fig:scans}a). However, this sensitivity has died away by about $M_H = 70$~GeV, since both the branching ratio to photons and the production cross section fall rapidly with increasing mass.
Other signatures also show potential for future sensitivity in high luminosity LHC data-taking. 

Fig.~\ref{fig:pools_a2}
shows a similar breakdown of the relevant final states for different values of $a_2$ and $M_H$ with $x_0 = 5.4$~GeV. It can be seen that the 95\% exclusion from the four-lepton measurement around $M_H = 200$~GeV extends to $a_2 \approx 0.2$.
There are regions at low $a_2$ and $M_H$ and at $0.6 < a_2 < 0.9$ where photon measurements dominate the sensitivity (both the ATLAS diphoton measurement, and the $\gamma$+\MET{} measurement~\cite{ATLAS:2018nci}). This corresponds to the fermiophobic case discussed in Section~\ref{neutralhiggs}, but is partly masked by the granularity of the scan.
In Fig.~\ref{fig:pools_a2-high} we show a higher granularity scan where the structure can be more clearly seen, as the band of yellow points extending from around $M_H = 70$ GeV, $a_2 = 0.1$ to $M_H = 150$ GeV, $a_2 = 0.9$.

Fig.~\ref{fig:diphoton} shows the predicted impact on the diphoton  invariant mass measurement which leads to these exclusions, for $a_2 = 0.3$ and $M_H = 95$~GeV. While this signal is excluded, by choosing an appropriate value of $a_2$ in this region it would be possible to produce a
signal in the two-photon mass distribution consistent with either the marginal excess at $M_H = 95$~GeV reported
by CMS~\cite{CMS:2018cyk}, or any other mass on the edge of the sensitivity of current low-mass searches~\cite{CMS:2023yay,ATLAS:2014jdv,ATLAS:2023jzc}.
\section{SUMMARY}
\label{sec:Summary}
We have discussed the main properties of the Higgs bosons present in the simplest model predicting a breaking of custodial symmetry at tree level. The Higgs sector of this model contains the SM Higgs doublet and a real $SU(2)_L$ triplet.
Tension in recent $W$ mass measurements~\cite{Amoroso:2023pey} motivates us to seriously consider this model, since it predicts an increase in the $W$-mass without affecting the value of the $Z$-mass at tree level. The main production mechanism for the new charged and neutral Higgs is Drell-Yan production due to the fact that the new Higgs bosons have suppressed couplings to the SM fermions. Somewhat to our surprise, these new Higgs bosons can be very light and yet still be consistent with experimental and theoretical constraints. The only current strong exclusion of the theoretically allowed region is for masses below around 60~GeV coming from diphoton decays of the neutral Higgs, and a band around 200~GeV coming from four-lepton decays of the neutral Higgs via two $Z$ bosons. The latter exclusion only applies for the $a_2$ mixing parameter is below $0.2$. The channel with two tau leptons and two neutrinos provides a promising tool to look for the light charged Higgses in future measurements.

We discussed in detail the scenario where the new neutral Higgs is fermiophobic, and can therefore have a large branching ratio into photons and/or $W$ gauge bosons. We briefly investigate the possibility of producing a signal in the diphoton mass spectrum, and show that is easily possible depending
on the value of $a_2$. Extreme fermiophobic scenarios are already excluded at low Higgs masses by bounds on SM $h \rightarrow \gamma\gamma$, and by diphoton measurements and searches, but these latter exclusions only apply for a narrow band of values of $a_2$.
The decay $H^\pm \to Z W^\pm$ remains particularly important because it provides direct information about custodial symmetry breaking in this context, and in this model its branching ratio can be large in the intermediate mass region. In this article we provide the first detailed study of all signatures in this model and use the available LHC data to understand the allowed scenarios.
We hope this study motivates new searches, and especially precision cross-section measurements of diboson production and of $\tau$ production, at the LHC and its high-luminosity upgrade.

{\textit{Acknowledgments}}: The work of H.D. and P.F.P. is supported by the U.S. Department of Energy, Office of Science, Office of High Energy
Physics, under Award Number DE-SC0024160.
\appendix
\section{FEYNMAN RULES}
\begin{eqnarray}
 h f\bar{f}  & : & \hspace{0.5 cm} i(M_f/v_0)\cos\theta_0,  
  \\
  H f\bar{f}  & : &   \hspace{0.4 cm} -i(M_f/v_0)\sin\theta_0,
  \\
  H^+ \bar{u} d  & : & -i \frac{\sqrt{2}}{v_0} \sin{\theta_{+}}  \left( - m_u \ V_{CKM} \ P_L \ + \ V_{CKM} \ m_d  \ P_R \right), \\
 H^+H^- h &:& \hspace{0.5 cm} -i\big(- a_1 \cos{\theta_+}\sin{\theta_+}\cos{\theta_0}+\frac{1}{2}a_1\sin^2{\theta_+}\sin{\theta_0}+a_2v_0\cos^2{\theta_
+}\cos{\theta_0}\nonumber \\
&&\hspace{0.5 cm} + a_2x_0\sin^2{\theta_+}\sin{\theta_0} +2b_4x_0\cos^2{\theta_+}\sin{\theta_0}+2\lambda_0v_0\sin^2{\theta_+}\cos{\theta_0}\big),
\\
ZZ h &:& \hspace{0.5 cm}(2iM_Z^2/v_0)\cos{\theta_0} \ g^{\mu\nu},
 \\
 ZZ H &:&  \hspace{0.3 cm }-(2iM_Z^2/v_0) \sin{\theta_0} \ g^{\mu\nu},
 \\
  ZW^\pm H^\mp &:& \hspace{0.5 cm} ig_2\big(-g_2x_0\cos{\theta_+}\cos{\theta_w}+\frac{1}{2}g_1v_0\sin{\theta_+}\sin{\theta_w}\big) \ g^{\mu\nu},
  \\
  W^+W^- h &:& \hspace{0.5 cm} ig^2_2\big(\frac{1}{2}v_0\cos{\theta_0}+2x_0\sin{\theta_0}\big) \ g^{\mu\nu},
 \\
W^+W^- H &:& \hspace{0.5 cm} ig^2_2\big(-\frac{1}{2}v_0\sin{\theta_0}+2x_0\cos{\theta_0}\big) \ g^{\mu\nu},
 \\
\gamma H^+H^-&:& \hspace{0.5 cm} ie\,\big(p'-p\big)^\mu, 
 \\
 ZH^+H^- &:& \hspace{0.5 cm}i\big(g_2\cos{\theta_w}-\frac{M_Z}{v_0}\sin^2{\theta_+})\big(p'-p\big)^\mu,
 \\
W^\pm h H^\mp &:&  \hspace{0.5 cm }\pm ig_2\big(\frac{1}{2}\sin{\theta_+}\cos{\theta_0}+
\cos{\theta_+}\sin{\theta_0}\big)\big(p'-p\big)^\mu,
 \\
 W^\pm H H^\mp &:& \hspace{0.5 cm} \mp ig_2\big(\frac{1}{2}\sin{\theta_+}\sin{\theta_0}-\cos{\theta_+}\cos{\theta_0}\big)\big(p'-p
\big)^\mu.
\end{eqnarray}

\section{Vacuum Structure}
\label{appendix2}
The value of the scalar potential for non-zero $x_0$ and $v_0$ reads as:
\begin{equation}
    V(v_0,x_0)= - \mu^2 \frac{v_0^2}{2} + \lambda_0 \frac{v_0^4}{4} - \frac{1}{2} M_{\Sigma}^2 x_0^2 + \frac{b_4}{4} x_0^4 - \frac{a_1}{4} v_0^2 x_0 + \frac{a_2}{4} v_0^2 x_0^2,
\end{equation}
One can satisfy these minimization conditions in four different cases: 1) $v_0=0$ and $x_0=0$, 2) $v_0=0$ and $x_0 \neq 0$, 3) $v_0 \neq 0$ and $x_0=0$, and 4) $v_0\neq 0$ and $x_0\neq 0$.
Notice that cases 1) and 2) are not phenomenological allowed because the electroweak symmetry is not broken. In case 3) one cannot explain a possible shift on the $W$-mass. The case 4) is the only scenario where one can break the electroweak symmetry and explain a possible shift on the $W$-mass mass. 

Now, we can study the different cases imposing the condition of minima, i.e.
\begin{equation}
\frac{\partial^2 V}{ \partial v_0^2} \frac{\partial^2 V}{ \partial x_0^2} - \left(\frac{\partial^2 V}{ \partial v_0 \partial x_0}\right)^2 > 0.
\label{conditiondet}
\end{equation}
and find conditions on the different free parameters.

\begin{itemize}
\item Case 1) $v_0=0$ and $x_0=0$: In this case the condition in Eq.(\ref{conditiondet}) requires $\mu^2 M_\Sigma^2 > 0$ and the energy of this vacuum is $V_1=0$.
\item Case 2) $v_0=0$ and $x_0 \neq 0$: In this case the condition in Eq.(\ref{conditiondet}) requires:
\begin{equation}
    \left(-\mu^2 -\frac{a_1}{2} \sqrt{\frac{M_\Sigma^2}{b_4}}+ \frac{a_2}{2} \frac{M_\Sigma^2}{b_4} \right) M_\Sigma^2 > 0.
\end{equation}
and the energy of this vacuum is given by
\begin{equation}
    V_2 = - \frac{1}{4} \frac{M_\Sigma^4}{b_4}=-\frac{1}{4} M_\Sigma^2 x_0^2.
\end{equation}

\item Case 3) $v_0 \neq 0$ and $x_0=0$: This case is only possible when $a_1=0$ (or when we have the symmetry $\Sigma \to - \Sigma$). The condition in Eq.(\ref{conditiondet}) requires that 
\begin{equation}
    \lambda_0 v_0^2 \left( - M_\Sigma^2 + \frac{a_2}{2} \frac{\mu^2}{\lambda_0}\right) > 0,
\end{equation}
while the energy of this vacuum reads as
\begin{equation}
V_3=-\frac{1}{4} \frac{\mu^4}{\lambda_0}=-\frac{1}{4} \lambda_0 v_0^4.    
\end{equation}
\item Case 4) $v_0\neq 0$ and $x_0\neq 0$: In this case the condition in Eq.(\ref{conditiondet}) requires that $b_4$ satisfy the Eq.(\ref{b4}). The energy of the vacuum reads as
\begin{equation}
V_4=-\frac{1}{4} \lambda_0 v_0^4 - \frac{1}{2} M_\Sigma^2 x_0^2 + b_4 \frac{x_0^4}{4}.  
\label{V4}
\end{equation}
Notice that this vacuum can be the global minimum only when $M_\Sigma^2 > 0$ because in this case one can satisfy the condition: $V_4 < V_3, V_2, V_1$. Notice that $\lambda_0$ and $b_4$ are positive, see the discussion above. 

Using Eqs.(\ref{mina}) and (\ref{minb}) one finds
\begin{eqnarray}
    a_1&=& \frac{4 (M_\Sigma^2 x_0^2 - b_4 x_0^4 + \lambda_0 v_0^4 - v_0^2 \mu^2)}{v_0^2 x_0}, 
   \label{a1}
    \\
    a_2 &=& \frac{2 (2 M_\Sigma^2 x_0^2  - 2 b_4 x_0^4 + \lambda_0 v_0^4  - v_0^2 \mu^2 )}{v_0^2 x_0^2}.
    \label{a2}
\end{eqnarray}
Therefore, one finds
\begin{equation}
a_2=\frac{a_1}{2 x_0} + 2 \left( \frac{M_\Sigma^2 - b_4 x_0^2}{v_0^2}\right) = 2 \frac{M_{H^\pm}^2}{v_0^2} + 2 \frac{M_\Sigma^2}{v_0^2} - 2 b_4 \frac{x_0^2}{v_0^2}.
\end{equation}
Notice that since $M_\Sigma^2 > 0$ the $a_2$ parameter is always positive. 
The last term in the above equation can be neglected because $v_0 \gg x_0$. Using the perturbative bound, i.e. $a_2 \leq 6 \pi$ $(\lambda_3 \leq 4 \pi, \lambda_4 \leq 4 \pi)$, one finds
\begin{equation}
M_{H^\pm}^2 \lesssim 3 \pi v_0^2  \ {\rm{or}} \ M_{H^\pm} \lesssim 755 \ {\rm{GeV}}.  
\end{equation}
This is a perturbative upper bound on the charged Higgs mass valid at tree level and assuming that the vacuum with $x_0 \neq 0$ and $v_0 \neq 0$ is the global minimum. Notice that new neutral Higgs mass is very close to the charged Higgs mass. Therefore, the above upper bound is also valid for the mass of the new neutral Higgs.

\end{itemize}

\bibliography{refs}

\end{document}